\documentclass{article}

\usepackage[sort&compress]{natbib}
\usepackage{graphicx}
\usepackage{amsmath,amsfonts,amssymb}
\usepackage{subfig}
\usepackage{hyperref}
\usepackage{caption}
\usepackage{tikz}
\usetikzlibrary{arrows,automata,positioning}
\usetikzlibrary{decorations.markings}
\usetikzlibrary{calc}

\begin{document}

\title{Complexity of {\sc mldp}}
\author{Nancy A. Arellano-Arriaga \and Juli\'{a}n Molina \and S. Elisa Schaeffer  \and  Ada M. \'Alvarez-Socarr\'as \and  Iris A. Mart\'inez-Salazar}

\maketitle

\begin{abstract}
  We carry out an explicit examination of the NP-hardness of a
  bi-objective optimization problem to minimize distance and latency
  of a single-vehicle route designed to serve a set of client
  requests. In addition to being a Hamiltonian cycle the route is to
  minimize the traveled distance of the vehicle as well as the total
  waiting time of the clients along the route.  
\end{abstract}

\section{Introduction}
\label{intro}

\textsc{mldp} consists in finding a route to visit all clients,
starting from and returning to a fixed depot, while minimizing the
total latency at the clients and the total traveled distance. In this
work, we limit to a {\em single} vehicle with no capacity restrictions
that visits all of the clients in a single tour. The problem is a
combination of the Minimum Latency Problem ({\sc mlp})
\citep{lucena-1990,Blum-etal-1994,Garcia-etal-2002,chaudhuri-etal-2003,Nagarajan-2008}
and the Traveling Salesman Problem ({\sc tsp})
\citep{laporte-1992,gutin-etal-2006,Applegate-etal-2007,Hoffman-etal-2013}.

In this work, we establish the NP-hardness of the problem, so as not
to further contribute to the prevailing trend in related literature of
omitting this step entirely and assuming that anything that resembles
something NP-hard is also NP-hard, without any need for formal
examination.

\section{Formulation of {\sc mldp}}

Consider a set of vertices $V=\{v_0, v_1, v_2,\ldots, v_n\}$, where
$v_0$ is the {\em depot} and the other $n$ are {\em clients}. A
matrix $\mathbf{T}$ contains $\forall (v_i, v_j)$ the {\em travel time}
from $i$ to $j$, $t_{i,j} \geq 0$, whereas $t_{i,i}$ represent the
{\em service time} at client $i$; we set $t_{0, 0} = 0$ and allow
$t_{i, j} \neq t_{j, i}$ A {\em cost matrix} $\mathbf{C}$ is given by
\begin{equation}
c_{i,j}= t_{i,i} + t_{i,j}.
\label{cost}
\end{equation}

Inspired by \citet{Picard-etal-1978}, we use variables $x_{i,k}$ to
indicate that vertex $i$ is the $k$th vertex to visit along the route;
we refer to the set of these binary variables as $\mathbf{X}$.  The
travel distance from the depot to the first client is
\begin{equation}
\mathcal{T}_0 = \sum_{i=1}^n c_{0,i} x_{i,1},
\end{equation}
the travel distance from the last client to the depot is
\begin{equation}
\mathcal{T}_n = \sum_{i=1}^n c_{i,0} x_{i,n},
\end{equation}
for which the {\em total travel distance} is
\begin{equation}
\mathcal{T} = \mathcal{T}_0 + \mathcal{T}_n + \sum_{k=1}^{n-1} \sum_{i=1}^n \sum_{j=1}^n c_{i,j} f_{i,j}(k).
\label{travel}
\end{equation}
The clients wait for the agent to reach and serve all the previously
visited clients
\begin{equation}
\mathcal{W}_0 = n \sum_{i=1}^n c_{0,i} x_{i,1},
\end{equation}
so the total waiting time (i.e., {\em latency}) is
\begin{equation}
\mathcal{W} =  \mathcal{W}_0 + \sum_{k=1}^{n-1} \sum_{i=1}^n \sum_{j=1}^n (n-k) c_{i,j} f_{i,j}(k).
\label{wait}
\end{equation}
Returning from the last client to the depot is not considered part of
the total latency of the clients, although it is included in Equation
\eqref{travel} for the total travel time.

Solving {\sc mldp} requires finding an assignment to $\mathbf{X}$ that
minimizes the objectives (Equations \eqref{travel} and \eqref{wait})
with equal importance) and satisfies the restrictions of visiting each
client exactly once and visiting a client at each step of the route
(i.e., that the $x_{i, k}$ form a valid permutation of the vertex
set).

\section{NP-hardness of {\sc mldp}}

The goal of an optimization problem $O_\sigma$ is to find the best
solution according to a set of constraints that describes the
environment of the problem. Each optimization problem is associated to
a decision problem $D_{\sigma}$ which differs from $O_\sigma$ in the
sense that the objective is treated as a constraint.  Given a bound
$B$ for the constraint representing the objective of $O_\sigma$, the
answer to the decision problem is whether or not an assignment exists
that satisfies both the original constraints and the new constraint
with bound $B$. Hence the decision problem $D_{\sigma}$ has only two
outcomes: yes or no. If the decision problem $D_{\sigma}$ is proven
NP-complete, then the optimization problem $O_\sigma$ is proven
NP-hard \citep{Garey-etal-1990,papadimitriou-1994}.

A {\em complexity class} is defined by several parameters such as
non-determinism and restrictions on the amount of computational
resources available (namely time and space)
\citep{papadimitriou-1994}. NP stands for \textit{non-deterministic
  polynomial time} meaning that the class is defined by bounding the
execution time polynomially under non-deterministic computation: given
an input for a problem $D_{\sigma}$, an ``oracle'' can guess a correct
solution in polynomial time. To prove the inclusion of any problem
$D_{\sigma}$ in the class NP it suffices to demonstrate that any
given solution of $O_\sigma$ can be verified as an actual solution of
$O_\sigma$ in polynomial time. Both of the single-objective
optimization problems that are merged into {\sc mldp} are NP-hard
\citep{Afrati-etal-1986,papadimitriou-1994}. In this section we
demonstrate that also {\sc mldp}itself is NP-hard.

We first define a decision problem $D_{\text{\sc mldp}}$ associated to
{\sc mldp}, then show that $D_{\text{\sc mldp}}$ belongs to NP, and
finally establish that an efficient reduction from a known
NP-complete problem to $D_{\text{\sc mldp}}$ exists
\citep{Garey-etal-1990}.

The corresponding decision problem is the following: given the costs
(Equation \eqref{cost}), an upper bound $\mathcal{T} \leq\Theta$ to
Equation \eqref{travel}, and an upper bound $\mathcal{W} \leq \Omega$
to Equation \eqref{wait}, is there an assignment $\mathbf{X}$ that
satisfies all the original constraints as well as the upper bounds set
on the objectives?

When service times and travel times are both constant, the problem is
trivial, as any visit order will be optimal. When service times are a
constant but travel times are arbitrary non-negative values, the
problem is NP-hard, as it is now simply a \textsc{tsp} instance. The
case of constant travel times and arbitrary non-negative service times
is an \textsc{mlp} instance. In this section, we detail the case that
both the service and the travel times are arbitrary.

The general \textsc{mldp} consists in finding a route which visits all
clients, leaves from a established depot and returns to it. All this
while minimizing the total latency of all the clients and the total
traveled distance of an uncapacitated vehicle.  We assume this agent
takes an arbitrary time $t_{ij}$ to reach every client and that all
service times $s_{i}$ are also arbitrary.

We first establish that the feasibility of a given $\mathbf{X}$ can be
verified in polynomial-time: each $\mathbf{X}$ captures a permutation
that starts at the depot and the permutation can be efficiently
recovered from the assignment matrix $\mathbf{X}$ as a visit sequence
$v^{(0)}, v^{(1)}, \ldots, v^{(n)}$ where $v^{(0)}$ is always the
depot and the visits from $v^1$ to $v^{(n)}$ correspond to the
clients.

We need three accumulator variables, one to measure time along the
route, one to count the total travel time, and the third to count the
total latency. We also need an array of $n$ binary decision variables,
one per each client, to verify that each one is properly visited.  We
proceed in the visiting order, denoting the source by $i$ and the
destination by $j$ adding the value of $t_{i,j}$ to both the time
accumulator and the travel-time accumulator, then adding the service
time $t_{i, i}$ to the time accumulator, and then the current value of
the time accumulator to the latency accumulator, making the client $j$
as visited. If at the end of the permutation, all $n$ binary variables
are one, the travel-time accumulator respects its upper bound
$\Theta$, and the latency accumulator respects its upper bound
$\Omega$, the solution is feasible. As the verification is a
polynomial procedure, $D_{\text{\sc mldp}} \in$ NP.

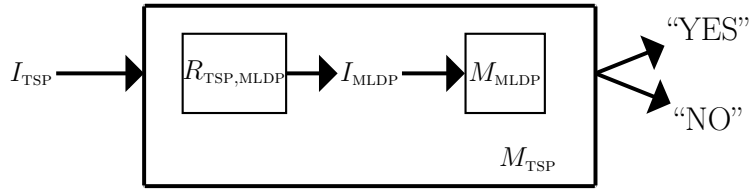
\begin{figure}[t]
     \centering
     \begin{tikzpicture}
       [client/.style={circle,draw=black!100,fill=black!100,thick,
inner sep=0pt,minimum size=0.1pt},
textos/.style={circle,draw=white!100,fill=white!100,thick,
inner sep=0pt,minimum size=10pt},
rect/.style={rectangle,draw=black!100,fill=white!100,thick,
inner sep=0pt,minimum size=50pt},
textos1/.style={circle,draw=black!100,fill=white!100,thick,
  inner sep=0pt,minimum size=10pt},]
 \begin{scope}[scale=0.6, transform shape]
\node[textos] (tetts) at (-2.5,2.5) {\LARGE{$I_\textsc{tsp}$}};
\node[client] (clienta) at (0,0) {};
\node[client] (clientb) at (10,0) {};
\node[client] (clientc) at (0,4) {};
\node[client] (clientd) at (10,4) {};
\node[rect] (rec1) at (2,2.5) {\LARGE{$R_{\textsc{tsp},\textsc{mldp}}$}};
\node[rect] (rec2) at (8,2.5) {\LARGE{$M_{\textsc{mldp}}$}};
\node[textos] (teets) at  (5,2.5) {\LARGE{$I_\textsc{mldp}$}};
\node[textos] (teets1) at  (12.5,1.5) {\huge{``NO''}};
\node[textos] (teets2) at  (12.5,3.5) {\huge{``YES''}};
\node[textos] (teets3) at (8.5,0.6) {\LARGE{$M_{\textsc{tsp}}$}};
\draw[-triangle 90,line width=1.5pt](tetts) to (0,2.5) {};
\draw[line width=1.5pt] (clienta) to (clientb) {};
\draw[line width=1.5pt] (clienta) to (clientc) {};
\draw[line width=1.5pt] (clientc) to (clientd) {};
\draw[line width=1.5pt] (clientb) to (clientd) {};
\draw[line width=1.5pt,-triangle 90,] (rec1) to (teets) {};
\draw[line width=1.5pt,-triangle 90,] (teets) to (rec2) {};
\draw[line width=1.5pt,-triangle 90,] (10,2.5) to (teets2) {};
\draw[line width=1.5pt,-triangle 90,] (10,2.5) to (teets1) {};
\end{scope}
     \end{tikzpicture}
     \caption{Reduction diagram from \textsc{tsp} to \textsc{mldp}
       (both as decision problems). The main box in the figure depicts
       an algorithm to solve the \textsc{tsp} problem that takes as
       input an instance of \textsc{tsp} and determines whether the
       answer is ``yes'' or ``no'' to this instance.}
    \label{reduct}
\end{figure}

To establish NP-completeness, we reduce the traveling salesman problem
(\textsc{tsp}) in its decision version to $D_{\text{\sc mldp}}$, as
illustrated in Figure \ref{reduct}. To prove the complexity of
\textsc{mldp}, a reduction algorithm must transform a \textsc{tsp}
entry into a \textsc{mldp} one. This \textsc{mldp} input is then
solved and receives a yes or no answer. As a whole, the process takes
a \textsc{tsp} instance, reduces it into an \textsc{mldp} instance,
solves the reduced instance with an algorithm for \textsc{mldp}, and
responds correctly ``yes'' or ``no'' to the original instance.

In the decision version of {\sc tsp}, the input is a cost matrix
$\mathbf{C}$ for the travel costs between $n$ clients (with the
diagonal elements being zero) together with an upper bound to the
total cost of the tour, $\mathcal{C}$, and the question is whether a
permutation over the set of clients exists that produces a tour with
the sum of costs of the segments not exceeding $\mathcal{C}$.

In order to transform the input for the decision version of the {\sc
  tsp} into an input for $D_{\text{\sc mldp}}$, we will use the
elements of $\mathbf{C}$ as $\mathcal{T}$, set $\Theta = \mathcal{C}$,
and only need to compute efficiently an adequate value for the latency
bound $\Omega$. We do this in terms of the worst-case costs, computing
from the cost matrix $\mathbf{C}$ the largest cost per each row (that
is, as if the agent at each client chose the most expensive segment to
continue the tour) in $\mathcal{O}(n^2)$ time, and then sort these
worst-case segments from largest to smallest in
$\mathcal{O}(n \log n)$ time to construct an upper bound to the
worst-case latency (the service times in {\sc tsp} are the diagonal
elements of the cost matrix that are all zero): the worst-case total
latency is the sum of the cumulative sums of the sorted list of costs
and we use this as $\Omega$. Hence, $D_{\text{\sc mldp}}$ responds
``yes'' to the transformed input if and only if the decision version
of {\sc tsp} would respond ``yes'' for the original input.

By reducing \textsc{tsp} to \textsc{mldp}, we show that \textsc{mldp}
is at least as difficult as \textsc{tsp}. Thus, $D_{\text{\sc mldp}}$
is NP-complete and therefore \textsc{mldp} is NP-hard.

\section{Conclusions}

We proved the NP-completeness of a decision problem corresponding to a
bi-objective problem that combines the Minimum Latency Problem ({\sc
  mlp}) and the Traveling Salesman Problem ({\sc tsp}).

%\bibliographystyle{plainnat}
%\bibliography{mldp}

\end{document}